\documentclass[12pt,preprint]{aastex}
\usepackage{epsfig}    
\usepackage[figuresright]{rotating}  
\usepackage{amsmath}   
\usepackage{amssymb}   
\usepackage{latexsym}  
\usepackage{dcolumn}
\newcolumntype{.}{D{.}{.}{-1}}

\begin{document}

\shorttitle{The Hanle Effect of the Ly-${\alpha}$ Line of He {\sc ii}}

\title{The Ly-${\alpha}$ Lines of H {\sc i} and He {\sc ii}: A Differential Hanle Effect for Exploring the Magnetism of the Solar Transition Region}

\author{Javier Trujillo Bueno\altaffilmark{1,2,3}, Ji\v{r}\'i \v{S}t\v{e}p\'an\altaffilmark{1,2,\dag} and Luca Belluzzi\altaffilmark{1,2}}
\altaffiltext{1}{Instituto de Astrof\'{\i}sica de Canarias, 38205 La Laguna,
Tenerife, Spain}
\altaffiltext{2}{Departamento de Astrof\'\i sica, Universidad de La Laguna, Tenerife, Spain}
\altaffiltext{3}{Consejo Superior de Investigaciones Cient\'{\i}ficas (Spain)}
\altaffiltext{\dag} {Associate Scientist at Astronomical Institute ASCR, Ond\v{r}ejov, Czech Republic}
\email{jtb@iac.es,stepan@iac.es,belluzzi@iac.es}

\begin{abstract}

The Ly-${\alpha}$ line of He {\sc ii} at 304 \AA\ is one of the spectral lines of choice for EUV channels of narrow-band imagers on board space telescopes, which provide spectacular intensity images of the outer solar atmosphere. Since the magnetic field information is encoded in the polarization of the spectral line radiation, it is important to investigate whether the He {\sc ii} line radiation from the solar disk can be polarized, along with its magnetic sensitivity. Here we report some theoretical predictions concerning the linear polarization signals produced by scattering processes in this strong emission line of the solar transition region, taking into account radiative transfer and the Hanle effect caused by the presence of organized and random magnetic fields. We find that the fractional polarization amplitudes are significant (${\sim}1\%$), even when considering the wavelength-integrated signals. Interestingly, the scattering polarization of the Ly-${\alpha}$ line of He {\sc ii} starts to be sensitive to the Hanle effect for magnetic strengths $B{\gtrsim}100$ G (i.e., for magnetic strengths of the order of and larger than the Hanle saturation field of the hydrogen Ly-${\alpha}$ line at 1216 \AA). We therefore propose simultaneous observations of the scattering polarization in both Ly-${\alpha}$ lines to facilitate magnetic field measurements in the upper solar chromosphere. Even the development of a narrow-band imaging polarimeter for the He {\sc ii} 304 \AA\ line alone would be already of great diagnostic value for probing the solar transition region. 

\end{abstract}

\keywords{Sun: magnetism --- Sun: chromosphere --- Sun: transition region --- radiative transfer --- polarization --- stars: atmospheres}

\section{Introduction}

The understanding of the complex interface region between the photosphere and corona of the Sun is a very important 
challenge in astrophysics. In this highly structured and dynamic region of the outer solar atmosphere, where magnetic and hydrodynamic forces compete for dominance, most of the non-thermal energy that creates the corona and solar wind is released. Novel measurements of key physical quantities, like the magnetic field, would improve our understanding of this boundary region. Spectroscopic observations are needed for determining temperatures, flows and waves, while the magnetic field information is encoded in the spectral line polarization (e.g., Stenflo 1994; Landi Degl'Innocenti \& Landolfi 2004). The aim of this paper is to propose a new technique, based on spectro-polarimetric measurements, which may be particularly useful for determining the magnetic field vector in the solar transition region.

The spectral lines that originate in the outer solar atmosphere (chromosphere, transition region and corona) 
are mainly located in the FUV (912-3000 \AA) and EUV (100-912 \AA) spectral ranges, such as the Ly-${\alpha}$ lines of H {\sc i} and He {\sc ii} at 1216 \AA\ and 304 \AA, respectively. Their intensity $I(\lambda)$ profiles are practically insensitive to the magnetic fields one may expect in the plasma of the outer solar atmosphere, so they cannot be used to obtain quantitative information on the strength ($B$), inclination ($\theta_B$) and azimuth ($\chi_B$) of the magnetic field vector. A similar drawback applies to the circular polarization (quantified by the Stokes $V(\lambda)$ profile) produced by the longitudinal Zeeman effect, because the Stokes-$V$ amplitude scales with the ratio, ${\cal R}$, between the Zeeman splitting and the Doppler line width. For such solar spectral lines ${\cal R}{\ll}1$, especially outside sunspots (${\cal R}={{1.4{\times}10^{-7}{\lambda}B}{/}{\sqrt{1.663{\times}10^{-2}T/\alpha+{\xi}^2}}}$, where $\lambda$ is the spectral line wavelength in \AA, $T$ the kinetic temperature in K, $\xi$ the microturbulent velocity in ${\rm km}\,{\rm s}^{-1}$, $\alpha$ the atomic weight of the atom under consideration, and $B$ is in gauss; see Landi Degl'Innocenti \& Landolfi 2004). The situation is even worse for the linear polarization (quantified by the Stokes $Q(\lambda)$ and $U(\lambda)$ profiles) produced by the transverse Zeeman effect because their amplitude scales with ${\cal R}^2$.

In order to obtain quantitative information on the magnetic field of the outer solar atmosphere we need to measure the polarization caused by scattering processes in the spectral lines that form in such regions, ideally using two or more spectral lines with different sensitivities to the Hanle effect (e.g., the review by Trujillo Bueno 2010; see also 
Stenflo, Keller \& Gandorfer 1998; Manso Sainz, Landi Degl'Innocenti \& Trujillo Bueno 2004). We recall that the Hanle effect is the modification of the linear polarization produced by scattering processes in a spectral line, caused by the presence of a magnetic field inclined with respect to the symmetry axis of the incident radiation field. Approximately, the emergent linear polarization is sensitive to magnetic strengths between 0.2$B_H$ and 5$B_H$, where $B_H{=}\,{1.137{\times}10^{-7}}/{t_{\rm life}g}$ is the critical Hanle field for which the Zeeman splitting of the line's level under consideration is equal to its natural width ($t_{\rm life}$ is the level's radiative lifetime, in seconds, and $g$ its Land\'e factor).  
  
In a recent paper we showed that the hydrogen Ly-${\alpha}$ line is expected to show measurable scattering polarization when observing the solar disk and that via the Hanle effect the line-center polarization amplitude is  sensitive to the presence of magnetic fields in the solar transition region, with good sensitivity to magnetic strengths between 10~G and 100~G (see Trujillo Bueno, \v{S}t\v{e}p\'an \& Casini 2011). The observational signatures of the Hanle effect might however be confused with the symmetry breaking effects caused by the presence of horizontal atmospheric inhomogeneities (e.g., Manso Sainz \& Trujillo Bueno 2011). Although for the hydrogen Ly-${\alpha}$ line such symmetry breaking effects can often be distinguished from the Hanle effect (e.g., \v{S}t\v{e}p\'an \& Trujillo Bueno 2012), it is of great interest to find another transition region line with measurable scattering polarization but such that is practically immune to the weak magnetic fields expected for the quiet regions of the upper solar chromosphere ($B{<}100$ G).

The aim of this Letter is to show some theoretical predictions concerning the linear polarization produced by scattering processes in the Ly-${\alpha}$ line of He {\sc ii}, whose significant emission originates in the solar transition region. As we shall see below, the line-center fractional polarization signals are significant 
(${\sim}1\%$) and have a very interesting magnetic sensitivity, which lead us to argue that the development of a space-based instrument capable of obtaining high-resolution $I$, $Q/I$ and $U/I$ images in the He {\sc ii} Ly-${\alpha}$ line would be very useful to determine the three-dimensional magnetic structure of the solar transition region.

\section{Formulation of the Problem}

The critical magnetic field strength for the onset of the Hanle effect is 
$B_H{\approx}53$ G for the Ly-${\alpha}$ line of hydrogen and $B_H{\approx}850$ G for the Ly-${\alpha}$ line of He {\sc ii}. Both spectral lines result from two blended transitions between a lower level, $^2{\rm S}_{1/2}$, and two upper levels, $^2{\rm P}_{1/2}$ and $^2{\rm P}_{3/2}$. In both cases the only level that contributes to the emergent linear polarization is the upper level $^2{\rm P}_{3/2}$, with total angular momentum $j=3/2$, whose Land\'e factor is $g=4/3$ (e.g., Trujillo Bueno et al. 2011). The reason why $B_H$ is 16 times larger for the Ly-${\alpha}$ line of He {\sc ii} is because its Einstein coefficient for spontaneous emission is 16 times larger (i.e., because the radiative lifetime of its $^2{\rm P}_{3/2}$ level is 16 times smaller). As we shall confirm below, this sizable difference between the $B_H$ values of the two Ly-${\alpha}$ lines implies that in the magnetic strength regime where the scattering polarization of the Ly-${\alpha}$ line of hydrogen is sensitive to the Hanle effect (i.e., between 10 and 100 G) there is little Hanle effect in the Ly-${\alpha}$ line of He {\sc ii}. 

In order to investigate the impact of the Hanle effect on the linear polarization amplitudes produced by scattering processes in the Ly-${\alpha}$ line of He {\sc ii}, we have followed the same complete frequency redistribution (CRD) approach we applied for estimating the line-center scattering polarization signals of the Ly-${\alpha}$ line of hydrogen (see Section 2 in Trujillo Bueno et al. 2011). As clarified in that paper, the CRD theory is suitable for estimating the fractional linear polarization at the line center, which is where the Hanle effect operates. The other approximation we have used is to neglect quantum interference between the $^2{\rm P}_{1/2}$ and $^2{\rm P}_{3/2}$ upper levels, which allows us to apply the multilevel atom approach described in Section 7.2 of Landi Degl'Innocenti \& Landolfi (2004). This approximation is justified for the H {\sc i} and He {\sc ii} Ly-${\alpha}$ line cores because in both cases the fine-structure (FS) splitting between such two levels is smaller than the Doppler width of the line, and, at the same time, it is much larger than the level's natural width (see Belluzzi \& Trujillo Bueno 2011). Indeed, magnetic fields of the order of kG (much larger than those expected for the outer solar atmosphere) are needed for entering the incomplete Paschen-Back regime, where the effect of $J$-state interference is no more negligible on the line-core polarization. It is interesting to observe that the ratio between the FS splitting between the $^2{\rm P}_{1/2}$ and $^2{\rm P}_{3/2}$ levels and their natural width is the same in H {\sc i} and He {\sc ii}. This is because for hydrogenic atoms both the Einstein coefficient for spontaneous emission and the FS splitting are proportional to the fourth power of the nuclear charge (e.g., Cowan 1981). Indeed, the Einstein coefficient of the Ly-${\alpha}$ line, and the FS splitting between the two upper levels is 16 times larger in He {\sc ii} than in H {\sc i}. 

The radiative transfer calculations needed to estimate the scattering polarization amplitudes of the H {\sc i} and He {\sc ii} Ly-${\alpha}$ lines have been carried out using the quiet atmosphere model C of Fontenla, Avrett \& Loeser (1993; hereafter, FAL-C model); this is sufficient for demonstrating the main point of this letter. We have used the H {\sc i} and He {\sc ii} number densities given by this semi-empirical model (see Fig. 1). 

The He {\sc ii} atomic model we have used includes the lower level ($1s\,{}^2{\rm S}_{1/2}$) and the two upper levels ($2p\,{}^2{\rm P}_{1/2}$ and $2p\,{}^2{\rm P}_{3/2}$). We quantify the excitation state of each $j$-level by means of the multipolar components of the atomic density matrix, whose self-consistent values at each spatial grid point of the model atmosphere have to be obtained by solving jointly the statistical equilibrium equations and the Stokes-vector transfer equation for each of the allowed radiative transitions in the atomic model. To this end, we have applied the multi-level radiative transfer code outlined in Appendix A of \v{S}t\v{e}p\'an \& Trujillo Bueno (2011), which is based on accurate and efficient radiative transfer methods. 

In addition to the above-mentioned radiative transitions, we have taken into account also inelastic collisional transitions between the $1s$ and $2p$ terms, using the electron cross-section data given by Janev et al. (1987). In our previous investigations on scattering polarization in the hydrogen Ly-$\alpha$ line (\v{S}t\v{e}p\'an \& Trujillo Bueno 2011, Trujillo Bueno et al. 2011), we took into account also the dipolar transitions (due to long-range collisional interactions with protons and electrons) between the metastable level $2s\,{}^2{\rm S}_{1/2}$ and the $2p\,{}^2{\rm P}_{j}$ levels. We took them into account because such collisional transitions can in principle cause depolarization of the $2p\,{}^2{\rm P}_{3/2}$ level (e.g., Sahal-Br{\'e}chot et~al. 1996). However, at the plasma densities of the upper chromosphere and transition region of the Sun, such collisional transitions do not noticeably affect the emergent linear polarization of the Ly-$\alpha$ line of hydrogen. Given that the depolarizing collisional rates are of the same order of magnitude for hydrogen and helium (Zygelman \& Dalgarno 1987) and that the radiative lifetime of the $2p\,{}^2{\rm P}_{3/2}$ level of He\,{\sc ii} is 16-times smaller than that of hydrogen, any collisional depolarization would be substantially smaller in the helium case. For this reason we have neglected such collisions as well as the Stark broadening when solving the radiative transfer problem for the He~{\sc ii} 304 \AA\ line. 

The continuum emissivity at the He {\sc ii} Ly-$\alpha$ wavelength is dominated by recombination processes producing H {\sc i} and He {\sc i}. Likewise, the continuum opacity is mainly due to photoionizations. Below the Lyman limit at 912 \AA\ the contributions from Rayleigh and Thomson scattering are negligible compared with that from radiative ionizations (e.g., Stenflo 2005). In the FAL-C model atmosphere, the continuum opacity at 304\,\AA\ becomes significant below a height of about $1900$\,km, but at this height the local Planck function is already very small ($\sim 10^{-26}\,{\rm erg\,cm^{-2}\,s^{-1}\,srad^{-1}\,Hz^{-1}}$). As a result, below this height, the continuum emissivity at the wavelength of the Ly$\alpha$ line of He~{\sc ii} is negligible. The He\,{\sc ii} Ly-$\alpha$ line is therefore formed in a zone of the transition region below which we have an essentially dark chromosphere. In effect, we have found through numerical experiments that the continuum processes do not affect the line-center polarization signal of the Ly$\alpha$ line of He {\sc ii}. 

With the above-mentioned physical ingredients our non-LTE synthesis of the intensity profile of the Ly-${\alpha}$ line of He {\sc ii} is in excellent agreement with that given in figure 12 of Fontenla et al. (1993).

\section{The anisotropy of the Ly-${\alpha}$ line of He {\sc ii} in the solar transition region}

The fractional anisotropy of the spectral line radiation, ${{{\bar J}^2_0}/{{\bar J}^0_0}}$, is the fundamental quantity that determines the largest fractional linear polarization amplitude we may expect to find in the emergent spectral line radiation (see equation (1) in Trujillo Bueno et al. 2011). The magnitude and sign of ${{{\bar J}^2_0}/{{\bar J}^0_0}}$ is sensitive to the gradient of the Stokes-$I$ component of the source function, $S_I$ (e.g., Trujillo Bueno 2001). Figure 2 illustrates the behavior of the fractional anisotropy of the local radiation for three different source-function gradients in a grey model atmosphere.

The left panel of Fig. 3 shows the height variation of $S_I$ of the Ly-${\alpha}$ line of He {\sc ii} after obtaining the self-consistent solution of the ensuing radiative transfer problem in the FAL-C model atmosphere. The right panel of Fig. 3 gives the corresponding variation of the fractional anisotropy, distinguishing between the contributions of the outgoing and incoming radiation intensities. As in the hydrogen Ly-${\alpha}$ case the fractional anisotropy of the Ly-${\alpha}$ line of He {\sc ii} is practically zero all through the model atmosphere, except in the model's transition region where it is negative and significant (i.e., of the order of a few percent at the atmospheric heights where the line-center optical depth is unity along the line of sight). 

\section{Hanle effect in the Ly-${\alpha}$ lines of H {\sc i} and He {\sc ii}}

The Hanle effect operates mainly in the line core, which is precisely the line spectral region where the CRD theory we have applied provides a good estimation of the fractional linear polarization amplitudes. For this reason, we summarize the main point of this letter in the two panels of Fig. 4, which show the magnetic sensitivity of the line-center $Q/I$ signals of the Ly-${\alpha}$ lines for two cases for which $U/I=0$. 

The left panel of Fig. 4 corresponds to the case of a horizontal magnetic field with a random azimuth, assuming a line of sight (LOS) with $\mu={\rm cos}\,{\theta}=0.3$ ($\theta$ being the heliocentric angle). As seen in this close to the limb scattering geometry, already in the unmagnetized case the linear polarization amplitude in the Ly-${\alpha}$ line of He {\sc ii} is about a factor three larger than in the Ly-${\alpha}$ line of hydrogen. Moreover, note that in the magnetic field regime where the Ly-${\alpha}$ line of hydrogen is sensitive to the Hanle effect (i.e., between 10 and 100 G, approximately) there is no significant magnetic sensitivity in the Ly-${\alpha}$ line of He {\sc ii}. This spectral line is sensitive to the Hanle effect for magnetic strengths between 100 and 1000 G, approximately, which makes it also of interest for probing the transition region plasma in solar active regions.

The right panel of Fig. 4 shows the case of a horizontal field with a fixed azimuth, for a LOS with $\mu=1$ (forward scattering geometry). In this scattering geometry the Hanle effect creates linear polarization (e.g., Trujillo Bueno et al. 2002). The largest polarization amplitude is reached for magnetic strengths $B>100$ G for the Ly-${\alpha}$ line of hydrogen and for $B>1000$ G for the Ly-${\alpha}$ line of He {\sc ii}. Note that for magnetic strengths $B<100$ G the Ly-${\alpha}$ line of He {\sc ii} is practically immune to the Hanle effect.

Finally, in Figure 5 we show Hanle diagrams for a close to the limb scattering geometry ($\mu=0.3$) 
assuming a horizontal magnetic field with a given azimuth (i.e., cases for which both $Q/I$ and $U/I$ 
are non-zero, in general). While the left panel shows the $Q/I$ and $U/I$ line-center signals corresponding to increasingly larger magnetic strengths of a horizontal field with a fixed azimuth ($\chi_B=0^{\circ}$), the right panel considers all possible magnetic field azimuths for two fixed magnetic strengths (25 G and 400 G). 

\section{Concluding comments}

The Stokes $I(\lambda)$ profile of the Ly-${\alpha}$ line of He {\sc ii} at 304 \AA\ is about 10 times narrower than that of the Ly-${\alpha}$ line of H {\sc i} at 1216 \AA. While only the core of the hydrogen Ly-${\alpha}$ line originates in the solar transition region, the Ly-${\alpha}$ line of He {\sc ii} is emitted mostly in the solar transition region. Noteworthy is also that the observed line-center intensities in the Ly-${\alpha}$ lines of H {\sc i} and He {\sc ii} are of the same order of magnitude (e.g., Fontenla et al. 1993).

The radiative transfer investigation reported in this paper indicates that 
the line-center fractional linear polarization amplitude of the Ly-${\alpha}$ line of He {\sc ii} should be 
significantly larger than that expected for the Ly-${\alpha}$ line of H {\sc i} (e.g., a factor three larger at $\mu\,{\approx}\,0.3$ in the unmagnetized case). Moreover, the fractional linear polarization amplitude that results from the wavelength-integrated Stokes profiles of the He {\sc ii} line turns out to be similar to the line-center signal\footnote{In a forthcoming publication we will show that the fractional linear polarization amplitude obtained from the wavelength-integrated Stokes profiles is even larger when partial frequency redistribution effects are taken into account.}. These results partially compensate the fact that the total number of photons emerging from any solar disk position in the Ly-${\alpha}$ line of He {\sc ii} is significantly smaller than within a small (0.1 \AA) wavelength interval around the hydrogen Ly-${\alpha}$ line center. 

We have shown also that for magnetic strengths $B{<}100$ G the Ly-${\alpha}$ line of He {\sc ii} is nearly immune to magnetic fields\footnote{The fact that with a narrow-band instrument, designed 
to measure the intensity and polarization of the Ly-${\alpha}$ line of He {\sc ii}, we may have a small 
contribution from the nearby Si {\sc XI} emission at 303.3 \AA\ should not be a problem because the critical Hanle field of this line is also very large (i.e., $B_H{\approx}730$ G).}. This is particularly interesting because for magnetic strengths between 10 and 100 G the Ly-${\alpha}$ line of H {\sc i} is indeed sensitive to the Hanle effect. Therefore, outside active regions the Ly-${\alpha}$ line of He {\sc ii} can be used as a reasonable reference line for facilitating magnetic field ``measurements" via the Hanle effect in the Ly-${\alpha}$ line of H {\sc i}. 

All these results encourage the development of the following instruments for a space telescope: 

\begin{itemize}
\item A spectropolarimeter for measuring the line-core polarization of the Ly-${\alpha}$ line of H {\sc i} with a spectral resolution of at least 0.1 \AA.
\item A narrow-band polarimeter for obtaining intensity and linear polarization images of the solar transition region in the Ly-${\alpha}$ light of He {\sc ii}.
\end{itemize} 

Although the combined use of the two Ly-$\alpha$ lines opens up a new diagnostic window for magnetic-field measurements in the upper solar chromosphere, the interpretation of such spectro-polarimetric observations will still require radiative transfer calculations in realistic atmospheric models because the two spectral lines are not formed in exactly the same way (e.g., Jordan 1975; Fontenla, Avrett \& Loeser 2002; Pietarila \& Judge 2004).
In spite of such a complication the comparison between spectro-polarimetric observations in the two Ly-$\alpha$ lines can provide unique insights into the physics and geometry of the transition region.

Finally, it is of interest to note that off-limb observations of resonant scattering polarization in the He {\sc ii} 304 \AA\ line may be also useful for exploring the geometry and magnetic field structure of spicules, prominences and of 
the solar corona. 

{\bf Acknowledgments}
Financial support by the Spanish Ministry of Science and Innovation through projects AYA2010-18029 (Solar Magnetism and Astrophysical Spectropolarimetry) and CONSOLIDER INGENIO CSD 2009-00038 (Molecular Astrophysics: The Herschel and Alma Era) is gratefully acknowledged.


\newpage


\begin{figure}[t]
  \centering
\includegraphics[width=14.cm]{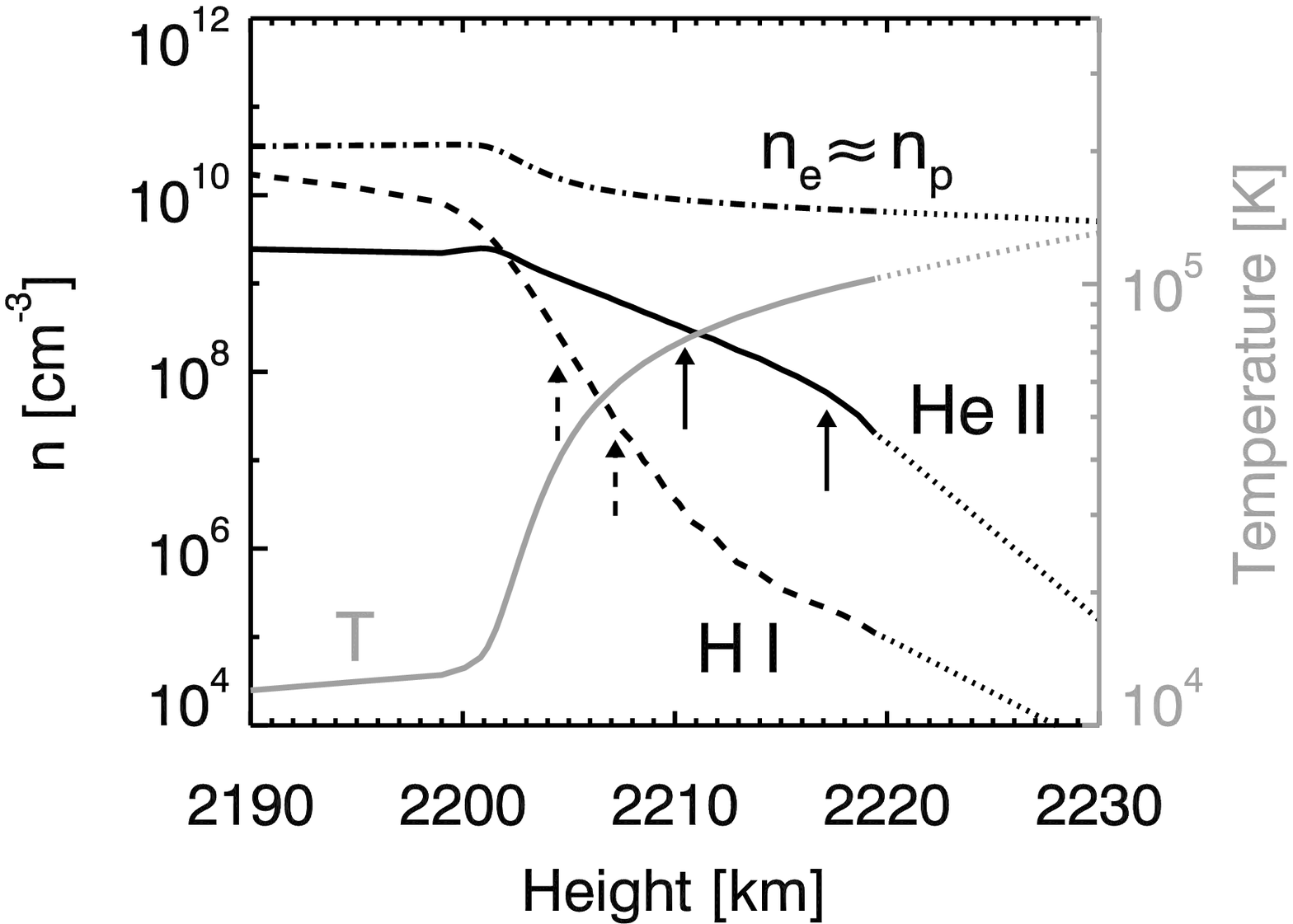}  
\caption[]{The transition region of the FAL-C model atmosphere, showing the 
height variation of the H {\sc i}, He {\sc ii} and electron number densities in addition to the 
kinetic temperature. The arrows indicate the heights where the line-center optical depth 
in both Ly-${\alpha}$ lines is unity for line of sights with $\mu=1$ (left arrow) and $\mu=0.1$ (right arrow). }
  
\label{fig:figure-1}
\end{figure}

\clearpage

\begin{figure}[t]
  \centering
\includegraphics[width=8.cm]{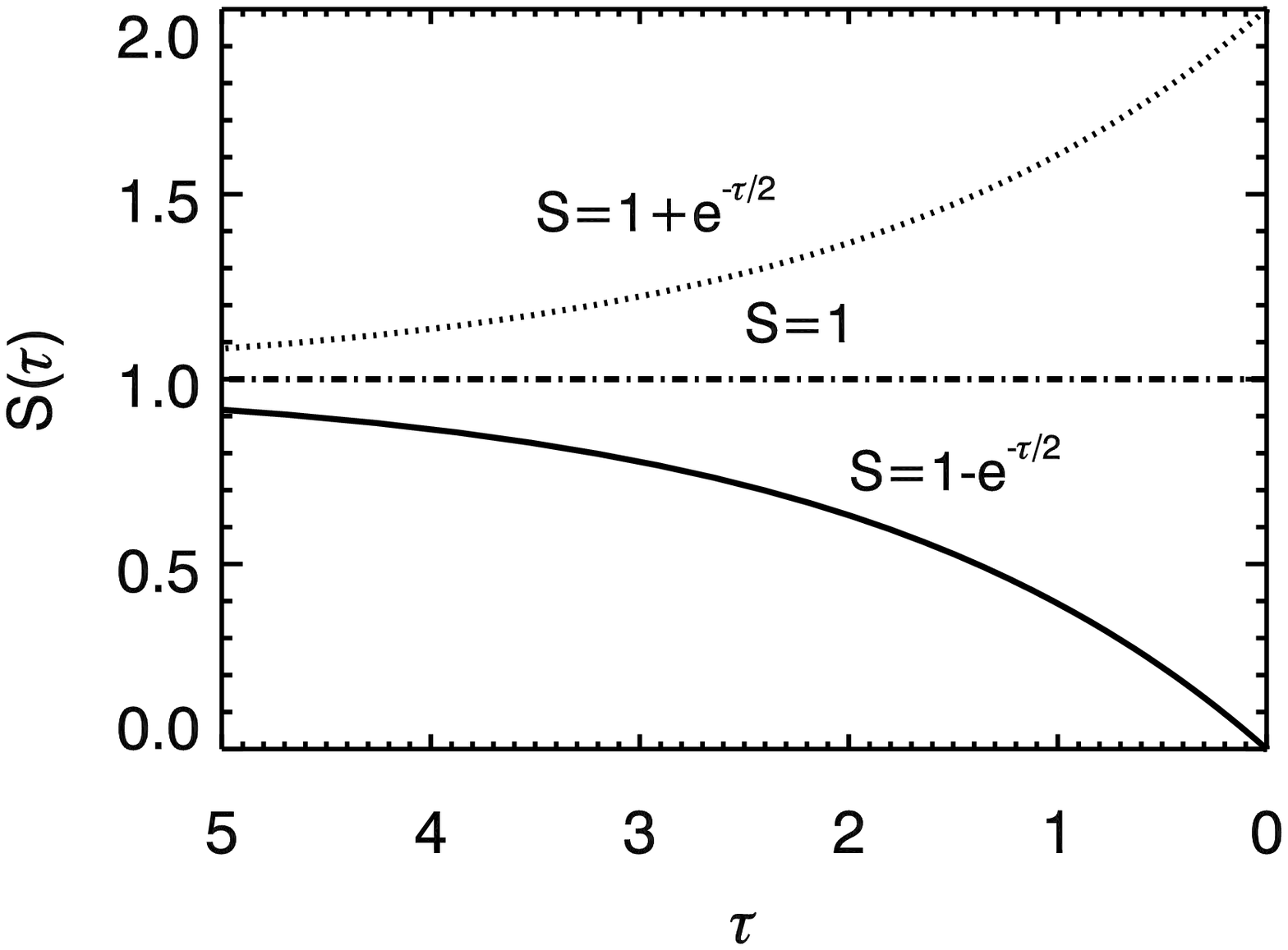}
\includegraphics[width=8.cm]{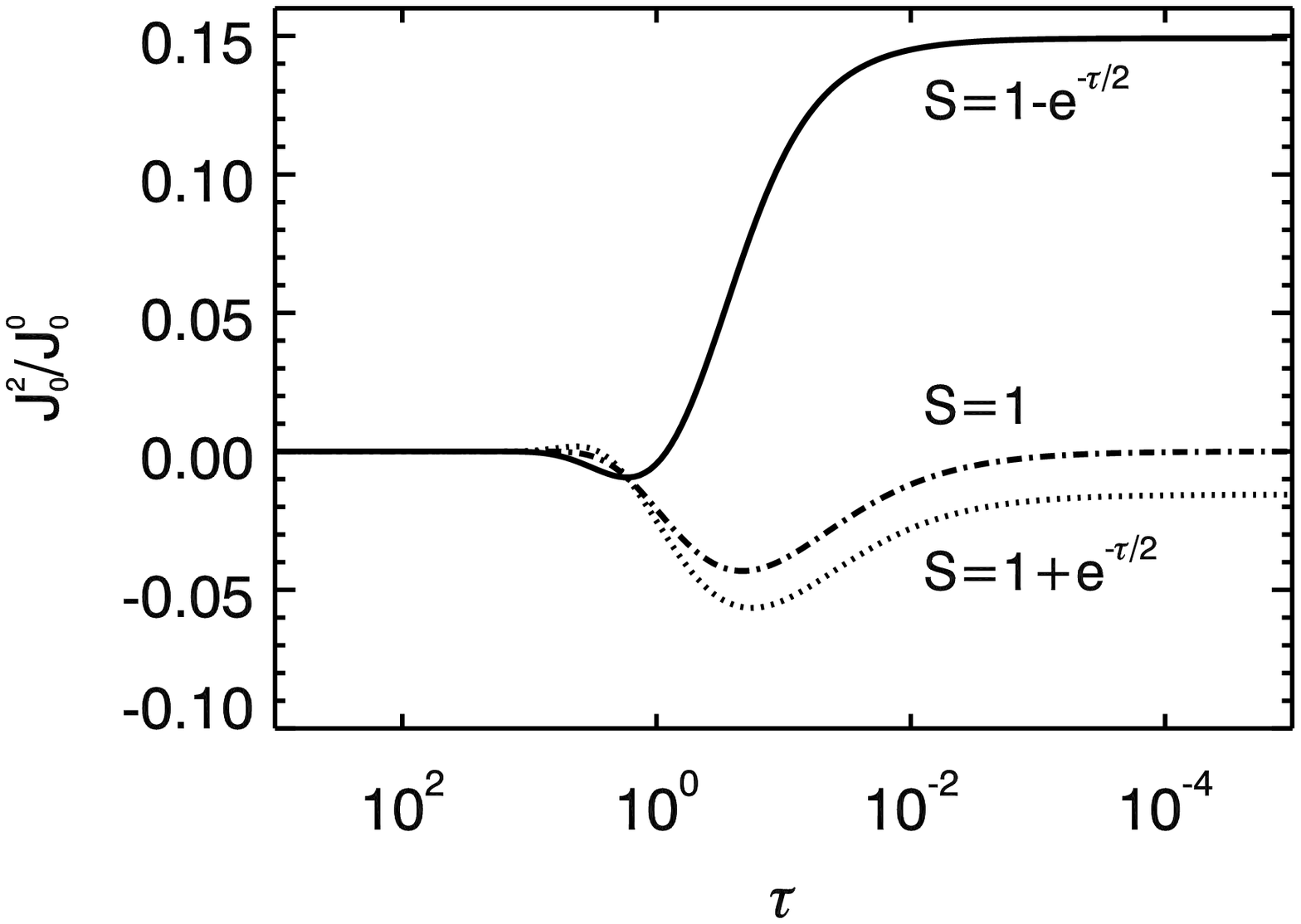}
  \caption[]{Illustration of the sensitivity of the fractional anisotropy of the local radiation field to the 
  gradient of the source function in a grey model atmosphere.}
\label{fig:figure-2}
\end{figure}

\clearpage 

\begin{figure}[t]
  \centering
\includegraphics[width=8.cm]{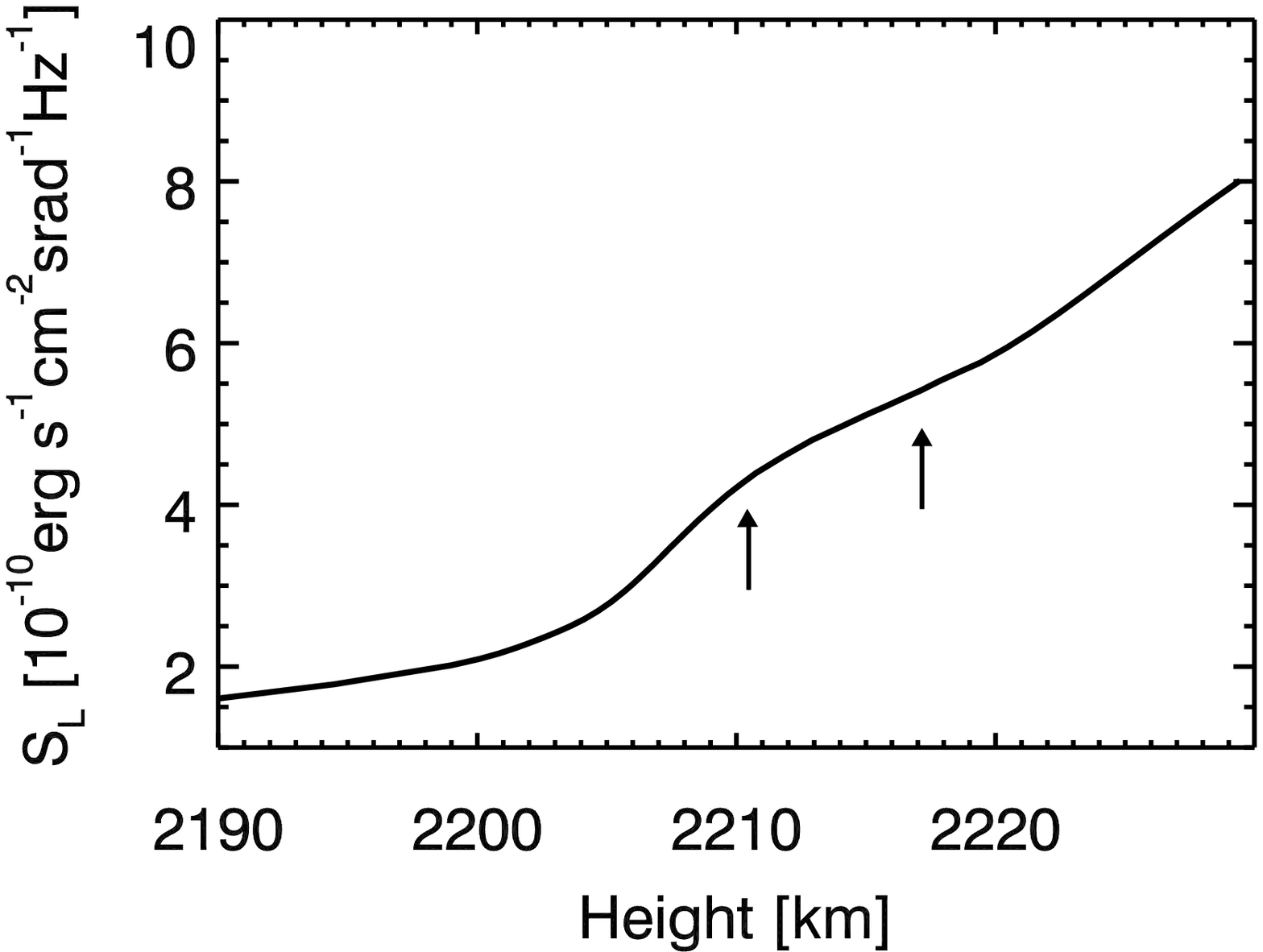}
\includegraphics[width=8.cm]{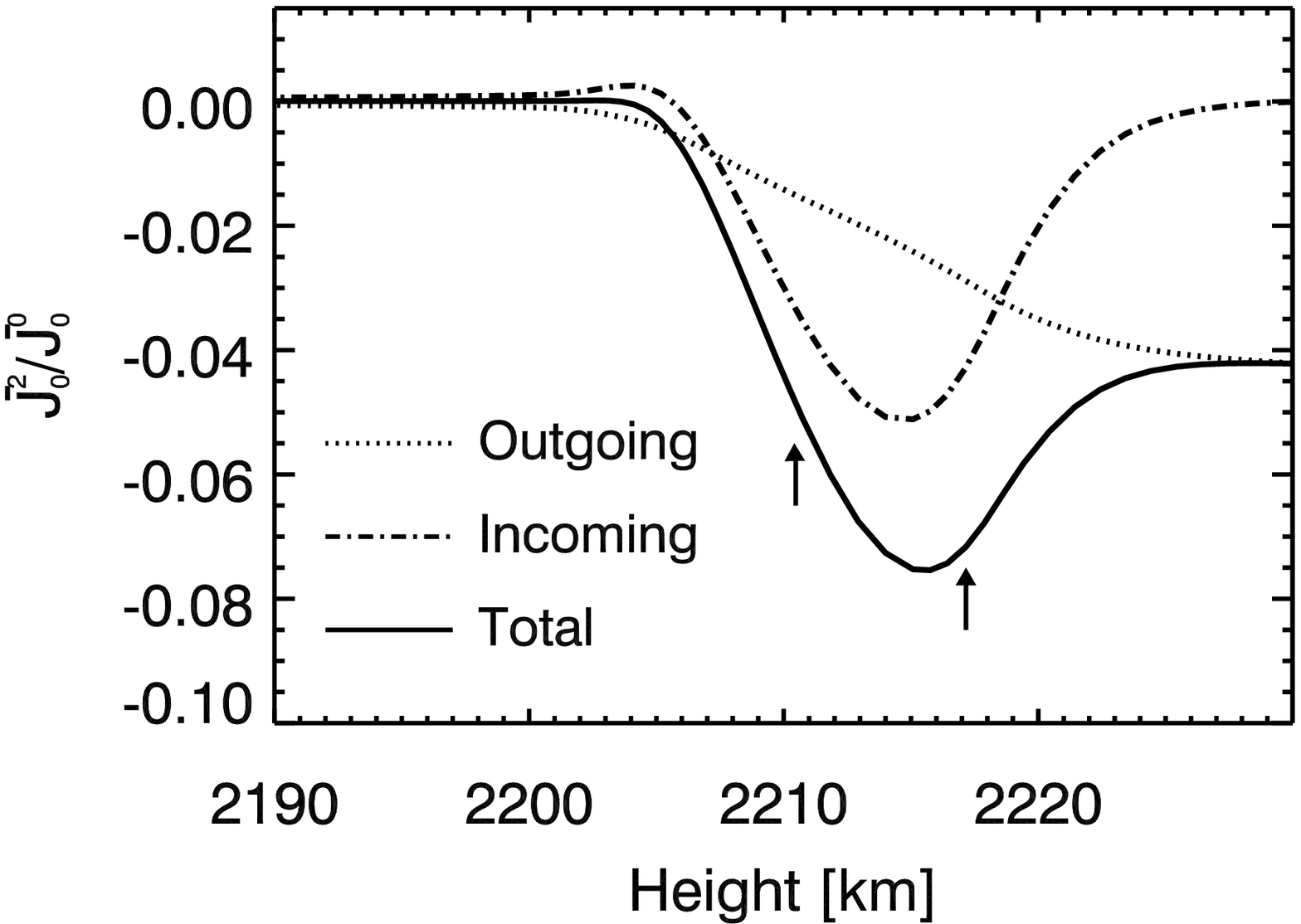}
  \caption[]{Source function and fractional anisotropy of the Ly-${\alpha}$ line of He {\sc ii}. Left panel: the variation with height in the FAL-C model atmosphere of the self-consistently calculated $I$-component of the source function of the Ly-${\alpha}$ line of He {\sc ii}.
  Right panel: the variation with height in the FAL-C model atmosphere of the fractional anisotropy of the Ly-${\alpha}$ line of He {\sc ii}, distinguishing between the contributions of the incoming radiation (with $-1{\le}{\mu}{<}0$) and outgoing radiation (with $0{<}{\mu}{\le}1$). The arrows indicate the heights where the line center optical depth is unity along line of sights with $\mu=1$ (left arrow) and $\mu=0.1$ (right arrow).}
\label{fig:figure-3}
\end{figure}

\clearpage

\begin{figure}[t]  
  \centering
\includegraphics[width=8cm]{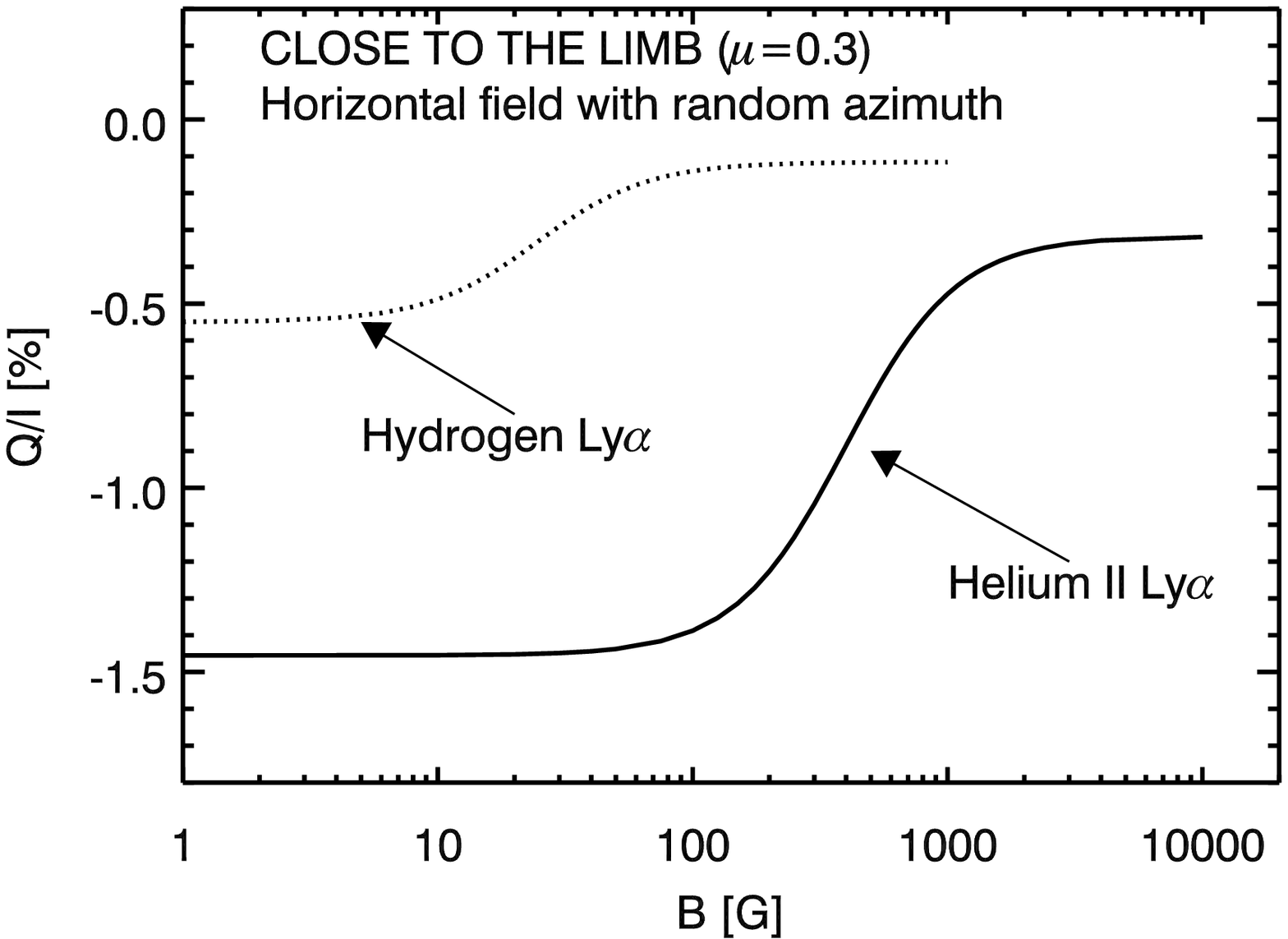}
\includegraphics[width=8cm]{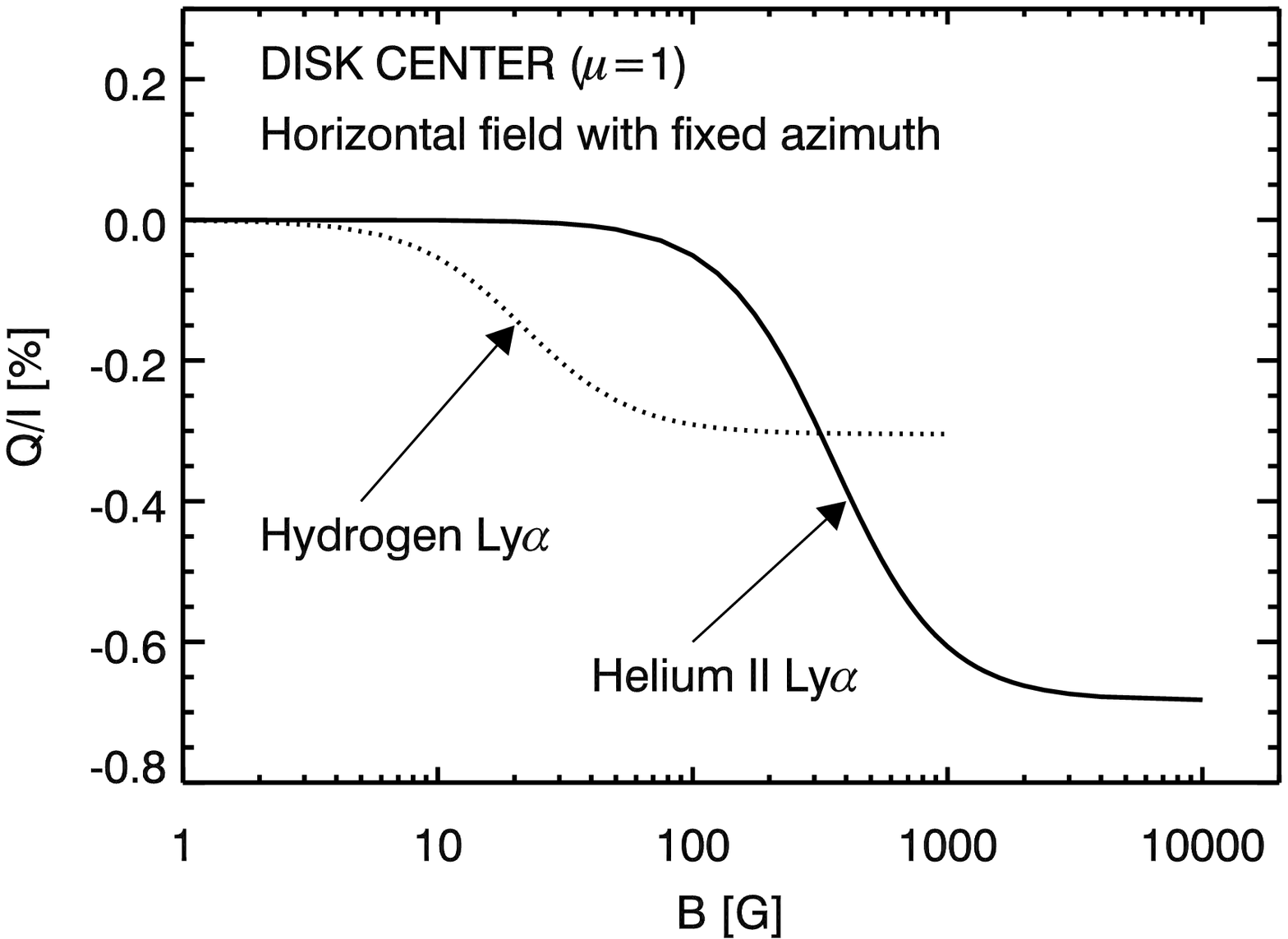}
  \caption[]{The sensitivity to the 
  Hanle effect of the scattering polarization amplitudes in the Ly-${\alpha}$ line of H {\sc i} (dotted lines) and in the Ly-${\alpha}$ line of He {\sc ii} (solid lines). Left panel: horizontal magnetic field with random azimuth for a 
  close to the limb scattering geometry (with $\mu=0.3$). The positive reference direction for Stokes $Q$ is the parallel to the nearest limb. 
  Right panel: horizontal field with fixed azimuth for the case of forward scattering geometry ($\mu=1$). The positive reference direction for Stokes $Q$ is along the magnetic field. 
  Note that for magnetic strengths $B<100$ G the He {\sc ii} line is practically 
  immune to magnetic fields, while the H {\sc i} line is sensitive to the Hanle effect.}
\label{fig:figure-4}
\end{figure}

\clearpage

\begin{figure}[t]  
  \centering
\includegraphics[width=8cm]{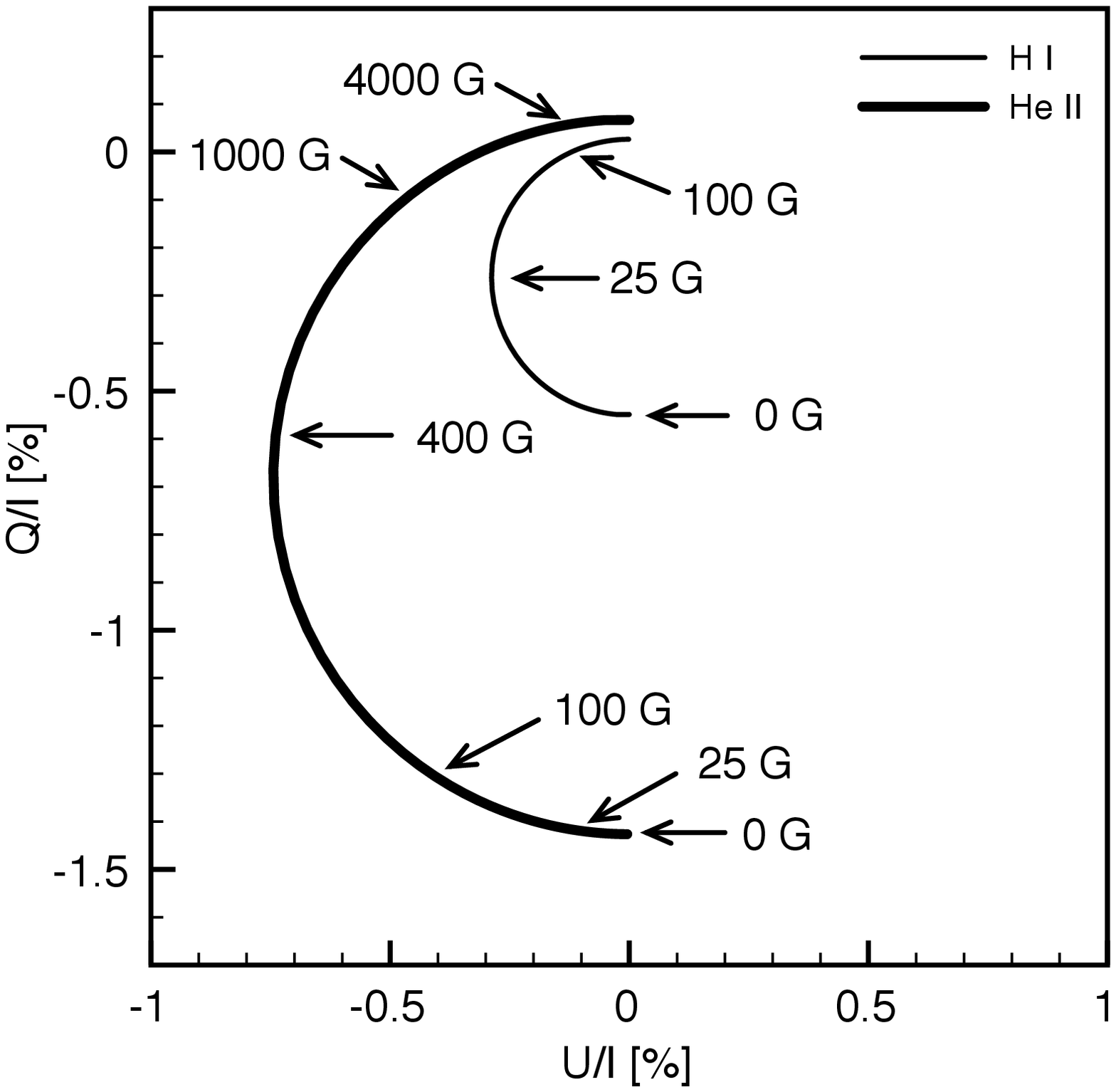}
\includegraphics[width=8cm]{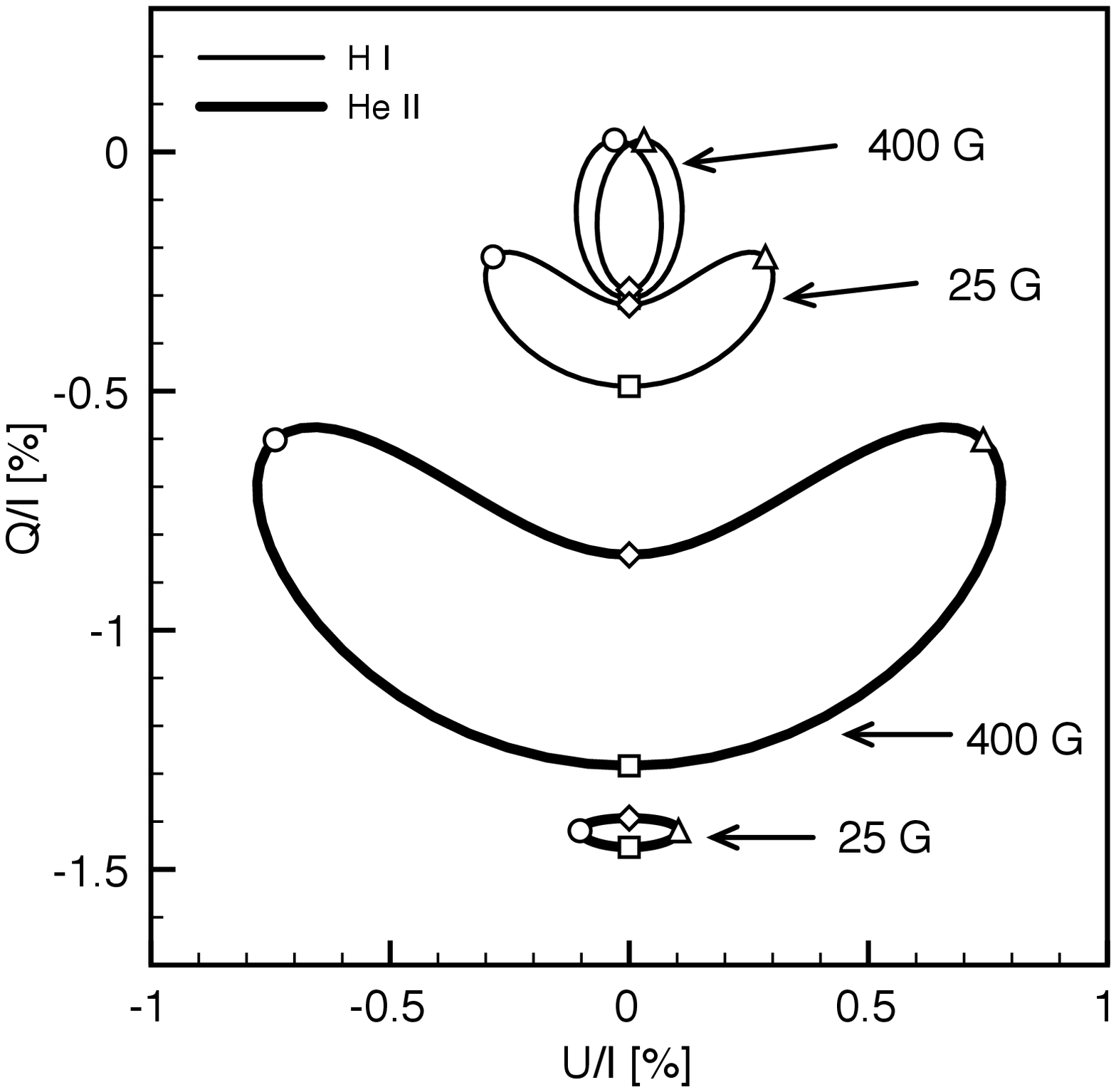}
\caption[]{Hanle diagrams of the Ly-${\alpha}$ lines of H {\sc i} and He {\sc ii} for a close to the limb scattering geometry ($\mu=0.3$). Left panel: $Q/I$ and $U/I$ line-center signals corresponding to increasingly larger magnetic strengths of a horizontal field with a fixed azimuth ($\chi_B=0^{\circ}$, with $\chi_B$ measured counterclockwise with respect to the projection of the LOS onto the solar surface plane). The right panel considers all possible magnetic field azimuths for the indicated magnetic strengths. Circles: $\chi_B=0^{\circ}$. Squares: $\chi_B=90^{\circ}$. Triangles: $\chi_B=180^{\circ}$. Diamonds: $\chi_B=270^{\circ}$. Note that for the 400 G case the hydrogen Ly-${\alpha}$ line is already in the Hanle saturation regime, where the $Q/I$ and $U/I$ signals are only sensitive to the orientation of the magnetic field.
}
\label{fig:figure-5}
\end{figure}

\end{document}